\documentclass[aip,rsi,reprint]{revtex4-1}

\usepackage{graphicx}
\usepackage{color}

\draft 

\begin{document}

\title{Compact Cryogenic Source of Periodic Hydrogen and Argon Droplet Beams for Relativistic Laser-Plasma Generation} 

\author{R. A. Costa Fraga}
\author{A. Kalinin}
\author{M. K\"uhnel}
\affiliation{Institut f\"ur Kernphysik, J. W. Goethe-Universit\"at, Max-von-Laue-Str. 1, 60438 Frankfurt am Main, Germany}
\author{D. C. Hochhaus}
\affiliation{EMMI Extreme Matter Institute and Research Division, GSI Helmholtzzentrum f\"ur Schwerionenforschung, Planckstr. 1, 64291 Darmstadt, Germany}
\affiliation{FIAS Frankfurt Institute for Advanced Studies, J. W. Goethe-Universit\"at, Ruth-Moufang-Str. 1, 60438 Frankfurt am Main, Germany}
\author{A. Schottelius}
\affiliation{Institut f\"ur Kernphysik, J. W. Goethe-Universit\"at, Max-von-Laue-Str. 1, 60438 Frankfurt am Main, Germany}
\author{J. Polz}
\affiliation{Institut f\"ur Optik und Quantenelektronik, Max-Wien-Platz 1, 07743 Jena, Germany}
\author{M. C. Kaluza}
\affiliation{Institut f\"ur Optik und Quantenelektronik, Max-Wien-Platz 1, 07743 Jena, Germany}
\affiliation{Helmholtz-Institut Jena, Fr\"obelstieg 3, 07743 Jena, Germany}
\author{P. Neumayer}
\affiliation{EMMI Extreme Matter Institute and Research Division, GSI Helmholtzzentrum f\"ur Schwerionenforschung, Planckstr. 1, 64291 Darmstadt, Germany}
\affiliation{FIAS Frankfurt Institute for Advanced Studies, J. W. Goethe-Universit\"at, Ruth-Moufang-Str. 1, 60438 Frankfurt am Main, Germany}
\author{R. E. Grisenti}
\email[]{grisenti@atom.uni-frankfurt.de}
\affiliation{Institut f\"ur Kernphysik, J. W. Goethe-Universit\"at, Max-von-Laue-Str. 1, 60438 Frankfurt am Main, Germany}
\affiliation{GSI Helmholtzzentrum f\"ur Schwerionenforschung, Planckstr. 1, 64291 Darmstadt, Germany}

\date{\today}

\begin{abstract}

We present a cryogenic source of periodic streams of micrometer-sized hydrogen and argon droplets as ideal mass-limited target systems for fundamental intense laser-driven plasma applications. The highly compact design combined with a high temporal and spatial droplet stability makes our injector ideally suited for experiments using state-of-the-art high-power lasers in which a precise synchronization between the laser pulses and the droplets is mandatory. We show this by irradiating argon droplets with multi-Terawatt pulses.

\end{abstract}


\maketitle 

\section{Introduction}
A liquid that is forced through a small orifice into vacuum under laminar flow conditions emerges as a continuous, cylindrical jet, before it eventually spontaneously breaks up into a stream of spherical droplets as a result of Rayleigh induced oscillations\cite{Eggers08}. Liquid jets deliver a uniquely functional, boundary-free and self-replenishing target beam, and have found widespread applications, e.g., for soft X-ray generation \cite{Malmqvist96}, X-ray absorption spectroscopy \cite{Smith04}, photoelectron spectroscopy \cite{Aziz08}, femtosecond X-ray crystallography \cite{Chapman11}, and also for studies of fast structural phase transformations \cite{Kuehnel11}.

Microscopic liquid jets are also very promising candidates for novel studies on intense laser-driven plasma generation. The interaction of ultrashort laser pulses with solid targets allows producing extreme conditions that are relevant to tabletop particle accelerators \cite{Hegelich06, Schwoerer06, Fuchs06} and laboratory astrophysics \cite{Glenzer09, Toleikis10}. Here, the laser energy is initially transferred to the target via the generation of relativistic electrons. However, the usually large dimensions of the employed targets, typically flat thin foils of mm$^2$ to cm$^2$ size, allow the hot electrons to spread transversely leading to a significant reduction of the energy density in the target. This precludes the efficient heating of the target material and thus has immediate consequences for fundamental applications such as ion acceleration\cite{Jaeckel10}. Recent efforts have provided evidence for very efficient bulk heating with the use of nearly mass-limited targets whose transverse dimensions are comparable to the laser focus such as microspheres \cite{Henig09} or microdots \cite{Neumayer09}, but the invariable presence of target holders still leads to a significant spreading of the electrons and hence to a rarefaction of the energy density. The use of levitated spherical targets in a Pauli trap has been demonstrated\cite{Sokollik10}, but these experiments still suffer from the major drawback that the employed target must be replaced after each laser shot. This, in turn, greatly precludes detailed parametric studies requiring the collection of a large amount of data, and the use of laser-driven ion sources operating in the (quasi-) continuous mode that is mandatory for most potential applications such as ion-based cancer therapy \cite{Kraft10}. Rayleigh droplet beams are extremely attractive with respect to the above applications because, when the Rayleigh instability is induced by an intentionally applied excitation, the ÒtriggeredÓ breakup process delivers a perfectly periodic stream of identical, isolated droplets at a production rate of up to $\sim 1$ MHz, thereby enabling detailed scaling studies under highly reproducible conditions by employing intense laser pulses in a wide range of repetition rates.

Experiments employing liquid water jets have clearly demonstrated the advantages of Rayleigh droplet beams for relativistic laser-plasma generation\cite{Ter-Avetisyan06, Sokollik09}, yet for many potential applications hydrogen and nobel gases represent the most scientifically relevant target systems. For example, it has been shown numerically that the use of a pure hydrogen target characterized by a higher plasma density would significantly increase the efficiency of the proton acceleration process\cite{Robinson09}, which would be further enhanced by the small droplet size. Hydrogen is also of central importance as model system for studies of the equation of state under high-density plasma conditions that are expected in the interior of giant planets such as Jupiter\cite{Toleikis10}. Liquid droplets of rare gases such as argon, on the other hand, are ideally suited for K-shell X-ray spectroscopy studies of the heating mechanisms of the bulk target material by providing direct access to the energy distribution and relaxation of the hot electron population\cite{Neumayer09}.

Whereas a variety of microscopic Rayleigh droplet beams, which include water, metallic and various organic solvent liquid jets, have been routinely produced in the laboratory for a decade\cite{Malmqvist96, Ostendal05, Weierstall08}, the stable generation of periodic droplet beams of cryogenic elements such as hydrogen and argon proves challenging. The high vapor pressure at the triple point of liquid argon and hydrogen results in very efficient evaporative cooling upon vacuum expansion. The expanding liquid filament thus rapidly cools below its normal melting point and freezes well before Rayleigh breakup can take place\cite{Hansson04, Kuehnel11}. The jet freezing can be circumvented by expanding the liquid into an atmosphere of the respective gas before injecting the resulting droplet stream into vacuum, a scheme that has enabled the production of hydrogen droplet beams for applications in nuclear physics research in a storage ring\cite{Nordhage05, Boukharov08}. However, the droplet sources employed in these studies are characterized by extended dimensions and a substantial loss of the spatial synchronization of the triggered droplet beam on the vacuum side\cite{Nordhage05}, crucial features that preclude their use in experiments in which pulses from a high-power laser are focused to a micro-scale spot. Here, we describe a novel concept for an injection source that addresses the above drawbacks delivering stable, periodic droplet beams of the cryogenic gases hydrogen and argon ideally suited for novel studies on relativistic laser-plasma generation.

\section{The cryogenic droplet injector}
At the heart of our droplet injector, shown in Fig. 1, is the use of a glass capillary that is inserted into an outer glass capillary tube (Fig. 1(b)). As the liquid jet emerges from the inner capillary it expands in an axially co-flowing gas plenum that suppresses evaporative cooling. The use of a co-flowing gas sheath has been demonstrated by Ga\~n\'an-Calvo as a method of generating columnar liquid jets of much reduced diameter as a results of gas dynamic forces\cite{Ganan-Calvo98}, and this approach has been adapted recently to liquid water jets\cite{DePonte08} as a means to deliver proteins into vacuum for serial crystallography\cite{Chapman11}. A novel and important aspect of the source described here is the additional challenge of working at cryogenic temperatures. In our design the typical distance between the inner capillary orifice and the outer tube exit hole is adjusted down to $\sim 1$ mm (Fig. 1(b)).  This feature thus allows significantly reducing the interaction time of the droplets with the co-flowing gas, which is held responsible for the degradation of the droplet spatial stability observed in previous studies\cite{Nordhage05}. The compact size of our source would greatly facilitate the droplet beam operation in a vacuum environment characterized by the presence of many delicate optical components as typically encountered in laser-plasma generation experiments.
\begin{figure}[t]
\includegraphics[width=1\linewidth]{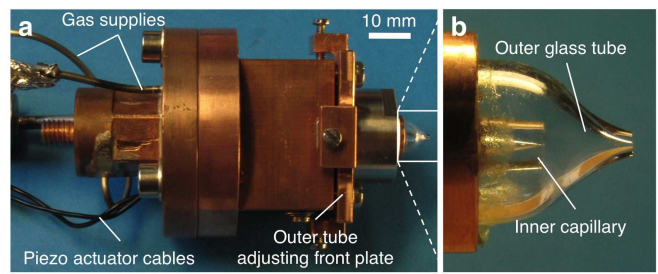}
\caption{\label{}(a) Picture of the compact cryogenic droplet injector (see the text). (b) Enlarged view of the exit end evidencing the outer glass tube, which has an exit hole diameter of 140 $\mu$m, and the central inner capillary.}
\end{figure}

The glass capillaries are produced from commercial, pulled Quartz tubes of $\approx 0.9$ mm inner diameter. Glass capillary nozzles allow producing liquid jets of much superior quality in terms of pointing stability as compared to commercially available thin-walled microscope apertures employed in previous studies\cite{Grisenti06}. We break the central neck under the microscope to precisely control the final orifice diameter. Our capillaries are characterized by a sharp orifice edge resulting in ideal conditions for laminar liquid flow. Each capillary is epoxy glued into a cylindrical copper plug, which is then sealed by compression onto a custom copper tailpiece. A ring-formed ultrasonic piezoelectric transducer that works at frequencies of up to 12.5 MHz is fixed to the capillary holder to generate a periodic disturbance that triggers the breakup of the liquid jet. Precise alignment of the outer glass tube exit aperture with respect to the jet propagation axis is obtained off-line by means of the adjusting front plate (Fig. 1(a)) while producing room-temperature isopropanol or water jets. During liquid beam operation in a vacuum environment our injector is screwed into the tip end of a continuous helium flow cryostat for cryogenic cooling of the source with a temperature stability of better than $0.02$ K.

Nozzle clogging is the most serious problem that arises when producing microscopic liquid jets from sub-10 $\mu$m diameter orifices. In general, by applying a moderate pressure from the capillary tip end it is possible to clean (i.e., Òde-clogÓ) a clogged nozzle, and we have found that the clog can be readily expelled from the glass tube in a centrifuge. Extensive observations with an optical microscope have evidenced that in our specific case the onset of clogging is primarily caused by the presence of sharp, a few $\mu$m long copper threads that protrude from the interior wall within the about 20 mm long channel downstream of a 0.5 $\mu$m pore filter. Our efforts devoted to significantly reduce the clogging rate have shown that a large fraction of these micro-scale threads, which are likely swept away by the flowing liquid, can be reliably removed by etching in acid (FeCl$_3$) solution the diverse copper tailpieces that constitute the nozzle assembly. This procedure allows us producing liquid jets from capillaries of diameter down to about 3 $\mu$m, and such beams would run continuously for several days. 

\section{Droplet beam operation}
Figures 2 shows periodic beams of monodisperse argon and hydrogen droplets of diameter of $21\pm 1$ $\mu$m and $13\pm 2$ $\mu$m, respectively, jetting from the outer tube exit hole into vacuum. Beam imaging is obtained by using a 4M pixels CCD camera and a long-distance microscope. The function generator driving the piezoelectric actuator also triggers a 10 Hz Nd-YAG laser emitting 10 ns pulses for stroboscopic backside illumination, as verified by observation of droplets that appear stationary at the piezo driving frequency. In particular, the stable, satellite-free argon and hydrogen droplet streams are found at excitation frequencies (within 2\%) $f=0.618$ MHz and 2.181 MHz, respectively. The argon and hydrogen droplets propagate at a velocity of $v=37\pm 1$ m$\,$s$^{-1}$ and $130\pm 3$ m$\,$s$^{-1}$, respectively, as inferred directly from the stroboscopic images according to $v=\lambda f$, where the distance between the droplets is $\lambda=59.6\pm 0.2$ $\mu$m for argon and $59.4\pm 0.2$ $\mu$m for hydrogen. The corresponding reduced wave number\cite{Eggers08} $x=\pi d/\lambda$ ($d$ is the orifice diameter, see caption of Fig. 2) is $\approx 0.53$ for the argon beam and $\approx 0.26$ for the hydrogen beam. These values are slightly different from that corresponding to the Rayleigh mode for fastest sinusoidal perturbation growth, which occurs for $x\approx 0.7$, and are consistent with previous observations\cite{Boukharov08}.
\begin{figure}[t]
\includegraphics[width=1\linewidth]{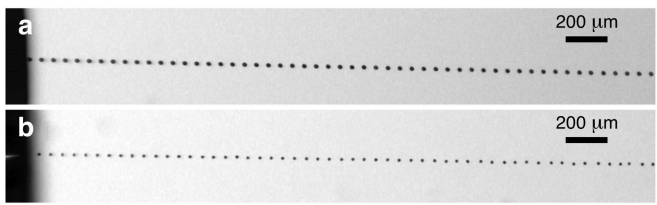}
\caption{\label{}Stroboscopic images of periodic (a) argon and (b) hydrogen droplet beams propagating in vacuum.  The argon beam is produced from a $10\pm 0.5$ $\mu$m diameter capillary orifice at a nominal stagnation source pressure of 10 bar and at a temperature of 86 K and expanding in argon gas at a mass flow of 35 sccm. The hydrogen beam is produced from a $ 5\pm 0.5$ $\mu$m diameter capillary orifice at a nominal stagnation source pressure of 7 bar and at a temperature of 14.5 K expanding in hydrogen gas at 70 sccm.}
\end{figure}

To show that our triggered droplet beams exhibit the necessary spatial stability we compare several single-shot images as those shown in Fig. 2. The individual frames are used to determine the relative radial (in the image plane) displacements of the center of mass positions of the individual droplets along the stream path\cite{Hemberg00}. The mean relative displacement $\left\langle\delta r\right\rangle$ obtained by averaging over all frames is plotted in Fig. 3 as a function of the distance $z$ from the outer tube exit aperture. For a more quantitative analysis of the droplet beam features of Fig. 3 we require a stability criterion that takes into account the laser spot radius $r_L$. A possible choice might be $\left\langle\delta r\right\rangle\leq r_L$, which would ensure a nearly ideal overlap between the laser pulse and the droplet. Thus, according to Fig. 3 we see that both hydrogen and argon droplet beams meet the above condition within the first few millimeters propagation distance for an experimentally realistic spot size radius of $r_L\geq 2$ $\mu$m.

The increase of $\left\langle\delta r\right\rangle$ with increasing $z$ (Fig. 3) results from the interaction of the droplets with the co-expanding gas prior to vacuum injection\cite{Nordhage05}, as evidenced in our case by the steeper increase of $\left\langle\delta r\right\rangle$ for the hydrogen droplets than for the much heavier argon droplets. This is further confirmed by the observation of spatially more stable hydrogen droplets produced without external gas atmosphere, as shown by the open symbols in Fig. 3. Here, the gas that evaporates from the hydrogen jet surface is trapped inside the outer glass tube, slowing down the freezing process and thus enabling the triggered jet breakup. However, the hydrogen droplet stream produced under these conditions is invariably characterized by the occasional appearance of solid, several tenths of micrometer long rods formed as a result of short-term jet freezing.
\begin{figure}[t]
\includegraphics[width=0.65\linewidth]{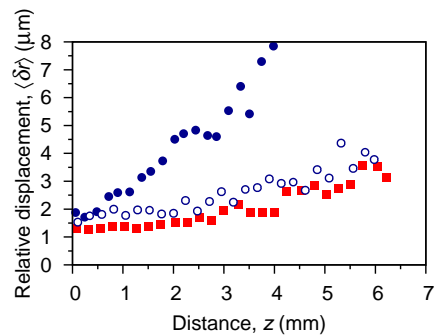}
\caption{\label{}Relative mean droplet displacement determined from several single-shot stroboscopic images plotted as a function of the distance from the outer tube exit hole. The filled symbols are for argon (squares) and hydrogen (circles) beams, respectively, expanding in a co-flowing gas. The open symbols are for a hydrogen droplet beam produced without external gas atmosphere.}
\end{figure}

\section{Relativistic laser-plasma generation}
In order to test our droplet injector, especially regarding the synchronization between the droplets and the laser pulses, we performed a proof-of-principle experiment at the PHELIX laser facility\cite{Bagnoud10} by employing the argon droplet beam as target. Laser pulses of 370 fs duration and containing 2.5 J energy from the pre-amplifier stage and generated at a repetition rate of about one shot every three minutes were focused by a $90^{\circ}$ off-axis parabola to a spot of $r_L \approx 3$ $\mu$m (see Fig. 4(a)) at a distance $z=4$ mm from the outer tube exit hole, producing peak intensities $\sim 10^{19}$ W cm$^{-2}$. Temporal synchronization was obtained by triggering the laser amplifier chain with the high-frequency signal that drives the Rayleigh breakup process. Since the short-pulse laser oscillator is not phase locked to the piezo frequency, the temporal accuracy was determined by the round-trip period of the pulses in the oscillator, resulting in a jitter of $\approx 13$ ns and corresponding to a spatial uncertainty of $\approx 0.45$ $\mu$m for the argon droplets. Thus, any deviation from the ideal case of a droplet uniformly illuminated by the laser pulse primarily came from the observed droplet relative spatial displacement at focus (Fig. 3).
\begin{figure}[t]
\includegraphics[width=0.75\linewidth]{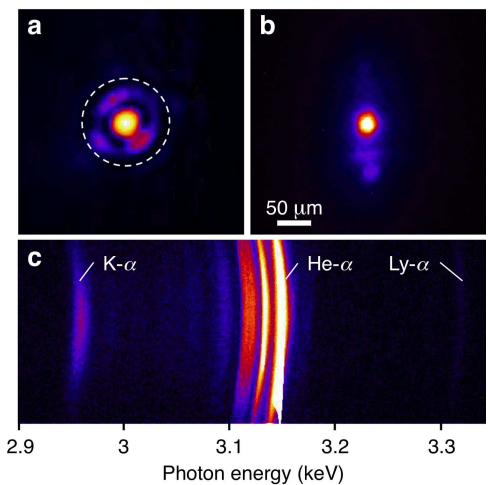}
\caption{\label{} (a) Focal laser intensity distribution. The dashed circle indicates the 21 $\mu$m diameter argon droplet. (b) Time-integrated VUV image of the droplet thermal emission. The laser pulse comes from the left.  (c) X-ray emission spectrum showing the strong argon K-$\alpha$ and He-$\alpha$ transition lines. The K-shell transitions in intermediate charge state argon ions are visible as satellites between the K-$\alpha$ and He-$\alpha$ lines.}
\end{figure}

Figure 4(b) shows a time-integrated 2D-image of the VUV emission in a narrow spectral range around 13.7 nm, evidencing a spherical plasma that remains close to the initial droplet size during the thermal emission stage, and thus supporting the conclusion that the droplets are heated isochorically to high-density plasma conditions. The droplet X-ray emission is spectrally dispersed by a cylindrically curved graphite crystal and is then recorded by a back-illuminated CCD camera. The measured X-ray spectrum is shown in Fig. 4(c), which covers the energy range from the argon cold K-$\alpha$ line at 2.96 keV to the Ly-$\alpha$ transition at 3.32 keV. The K-$\alpha$ line originates from the fluorescence decay following collisional ionization of K-shell electrons by the relativistic electron population produced during the interaction with the intense laser pulse. The estimated conversion efficiency of the laser energy into K-$\alpha$ radiation of $\sim 10^{-5}$ is in good agreement with previous experiments employing nearly mass-limited targets\cite{Neumayer09, Henig09}, indicating indeed an efficient coupling to the relativistic electrons.

\section{Conclusions}
Whereas a more detailed analysis of the recorded spectra will provide deeper insights into the heating mechanisms of the solid-density argon droplets, the results presented here clearly demonstrate the potential of our cryogenic droplet beam injector as a means to deliver ideal mass-limited target samples in vacuum. In particular, the compact design combined with a high spatial droplet stability makes the injection system described here ideally suited for relativistic laser-plasma applications in which a precise control on the overlap between the laser beam focus and the droplets is mandatory. Our droplet injector thus opens up new possibilities for tabletop proton accelerators and studies of matter under extreme conditions relevant to astrophysical phenomena.

\begin{acknowledgments}

This work was supported by the Helmholtz Gemeinschaft, through Grant No. VH-NG-331, and by the EC within the Seventh Framework Program (Project No. 227431).

\end{acknowledgments}



\begin{thebibliography}{00}

\bibitem{Eggers08}
J. Eggers and E. Villermaux, Rep. Prog. Phys. \textbf{71}, 036601 (2008).

\bibitem{Malmqvist96}
L. Malmqvist, L. Rymell, and H. M. Hertz, Appl. Phys. Lett. \textbf{68}, 2627 (1996).

\bibitem{Smith04}
J. D. Smith, C. D. Cappa, K. R. Wilson, B. M. Messer, R. C. Cohen, and R. J. Saykally, Science \textbf{306}, 851 (2004).

\bibitem{Aziz08}
E. F. Aziz, N. Ottosson, M. Faubel, I. V. Hertel, and B. Winter, Nature \textbf{455}, 89 (2008).


\bibitem{Chapman11}
H. N. Chapman, P. Fromme, A. Barty, T. A. White, R. A. Kirian, A. Aquila, M. S. Hunter, J. Schulz, D. P. DePonte, U. Weierstall,	 R. B. Doak, F. R. N. C. Maia, A. V. Martin, I. Schlichting, L. Lomb, N. Coppola, R. L. Shoeman, S. W. Epp,	 R. Hartmann, D. Rolles, A. Rudenko, L. Foucar, N. Kimmel, G. Weidenspointner, P. Holl,	 M. Liang, M. Barthelmess, C. Caleman, S. Boutet, M. J. Bogan, J. Krzywinski, C. Bostedt, S. Bajt,	L. Gumprecht, B. Rudek, B. Erk, C. Schmidt, A. H\"omke, C. Reich, D. Pietschner, L. Str\"uder, G. Hauser, H. Gorke, J. Ullrich, S. Herrmann, G. Schaller, F. Schopper, H. Soltau, K.-U. KŸhnel, M. Messerschmidt, J. D. Bozek, S. P. Hau-Riege, M. Frank, C. Y. Hampton, R. G. Sierra, D. Starodub, G. J. Williams, J. Hajdu, N. Timneanu, M. M. Seibert, J. Andreasson, A. Rocker, O. J\"onsson, M. Svenda, S. Stern, K. Nass, R. Andritschke, C.-D. Schr\"oter, F. Krasniqi, M. Bott, K. E. Schmidt, X. Wang, I. Grotjohann, J. M. Holton, T. R. M. Barends, R. Neutze, S. Marchesini, R. Fromme, S. Schorb, D. Rupp, M. Adolph, T. Gorkhover, I. Andersson, H. Hirsemann, G. Potdevin, H. Graafsma, B. Nilsson, and J. C. H. Spence, Nature \textbf{470}, 73 (2011).


\bibitem{Kuehnel11}
M. K\"uhnel, J. M. Fern\'andez, G. Tejeda, A. Kalinin, S. Montero, and R. E. Grisenti, Phys. Rev. Lett. \textbf{106}, 234501 (2011).




\bibitem{Hegelich06}
B. M. Hegelich, B. J. Albright, J. Cobble, K. Flippo, S. Letzring, M. Paffett, H. Ruhl, J. Schreiber, R. K. Schulze, and J. C. Fern\'andez, Nature \textbf{439}, 441 (2006).

\bibitem{Schwoerer06}
H. Schwoerer, S. Pfotenhauer, O. J\"ackel, K.-U. Amthor, B. Liesfeld, W. Ziegler, R. Sauerbrey, K. W. D. Ledingham, and T. Esirkepov, Nature \textbf{439}, 441 (2006).

\bibitem{Fuchs06}
J. Fuchs, P. Antici, E. D'Humi\`eres, E. Lefebvre, M. Borghesi, E. Brambrink, C. A. Cecchetti, M. Kaluza, V. Malka, M. Manclossi, S. Meyroneinc, P. Mora, J. Schreiber, T. Toncian, H. P\'epin, and P. Audebert, Nature Phys. \textbf{2}, 48 (2006).




\bibitem{Glenzer09}
S. H. Glenzer and R. Redmer, Rev. Mod. Phys. \textbf{81}, 1625 (2009).

\bibitem{Toleikis10}
S. Toleikis, Th. Bornath, T. D\"oppner, S. D\"usterer, R. R. F\"austlin, E. F\"orster, C. Fortmann, S. H. Glenzer, S. G\"ode, G. Gregori, R. Irsig, T. Laarmann, H. J. Lee, B. Li, K.-H. Meiwes-Broer, J. Mithen, B. Nagler, A. Przystawik, P. Radcliffe, H. Redlin, R. Redmer, H. Reinholz, G. R\"opke, F. Tavella, R. Thiele, J. Tiggesb\"aumker, I. Uschmann, S. M. Vinko, T. Whitcher, U. Zastrau, B. Ziaja, and Th. Tschentscher, J. Phys. B \textbf{43}, 194017 (2010).

\bibitem{Jaeckel10}
O. J\"ackel, J. Polz, S. M. Pfotenhauer, H.-P. Schlenvoigt, H. Schwoerer, and M. C. Kaluza, New J. Phys., \textbf{12}, 103027 (2010).



\bibitem{Henig09}
A. Henig, D. Kiefer, M. Geissler, S. G. Rykovanov, R. Ramis, R. H\"orlein, J. Osterhoff, Zs. Major, L. Veisz, S. Karsch, F. Krausz, D. Habs, and J. Schreiber, Phys. Rev. Lett. \textbf{102}, 095002 (2009).

\bibitem{Neumayer09}
P. Neumayer, H. J. Lee, D. Offerman, E. Shipton, A. Kemp, A. L. Kritcher, T. D\"oppner, C. A. Back, and S. H. Glenzer, High Energy Density Phys. \textbf{5}, 244 (2009).

\bibitem{Sokollik10}
T. Sokollik, T. Paasch-Colberg, K. Gorling, U. Eichmann, M. Schn\"urer, S. Steinke, P. V. Nickles, A. Andreev, and W. Sandner, New J. Phys., \textbf{12}, 113013 (2010).


\bibitem{Kraft10}
S. D. Kraft, C. Richter, K. Zeil, M. Baumann, E. Beyreuther, S. Bock, M. Bussmann, T. E. Cowan, Y. Dammene, W. Enghardt, U. Helbig, L. Karsch, T. Kluge, L. Laschinsky, E. Lessmann, J. Metzkes, D. Naumburger, R. Sauerbrey, M. Sch\"urer, M. Sobiella, J. Woithe, U. Schramm, and J. Pawelke, New. J. Phys. \textbf{12}, 085003 (2010).


\bibitem{Ter-Avetisyan06}
S. Ter-Avetisyan, M. Schn\"urer, P. V. Nickles, M. Kalashnikov, E. Risse, T. Sokollik, W. Sandner, A. Andreev, and V. Tikhonchuk, Phys. Rev. Lett. \textbf{96}, 145006 (2006).

\bibitem{Sokollik09}
T. Sokollik, M. Schn\"urer, S. Steinke, P. V. Nickles, W. Sandner, M. Amin, T. Toncian, O. Willi, and A. A. Andreev, Phys. Rev. Lett. \textbf{103}, 135003 (2009).

\bibitem{Robinson09}
A. P. L. Robinson, P. Gibbon, S. M. Pfotenhauer, O. J\"ackel, and J. Polz, Plasma Phys. Control. Fusion \textbf{51}, 024001 (2009).




\bibitem{Ostendal05}
M. Otendal, O. Hemberg, T. T. Tuohimaa, and H. M. Hertz, Exp. Fluids \textbf{39}, 799 (2005).

\bibitem{Weierstall08}
E. Weierstall, R. B. Doak, J. C. H. Spence, D. Starodub, D. Shapiro, P. Kennedy, J. Warner, G. G. Hembree, P. Fromme, and H. N. Chapman, Exp. Fluids \textbf{44}, 675 (2008).

\bibitem{Hansson04}
B. A. M. Hansson, M. Berglund, O. Hemberg, and H. M. Hertz, J. Appl. Phys. \textbf{95}, 4432 (2004).


\bibitem{Nordhage05}
\"O. Nordhage, Z.-K. Lib, C.-J. Frid\'en, G. Norman, and U. Wiedner, Nucl. Instr. Meth. A \textbf{546}, 391 (2005).

\bibitem{Boukharov08}
A. V. Boukharov, M. B\"uscher, A. S. Gerasimov, V. D. Chernetsky, P. V. Fedorets, I. N. Maryshev, A. A. Semenov, and A. F. Ginevskii, Phys. Rev. Lett. \textbf{100}, 174505 (2008).

\bibitem{Ganan-Calvo98}
A. M. Ga\~n\'an-Calvo, Phys. Rev. Lett. \textbf{85}, 285 (1998).

\bibitem{DePonte08}
D. P. DePonte, U. Weierstall, K. Schmidt, J. Warner, D. Starodub, J. C. H. Spence, and R. B. Doak, J. Phys. D \textbf{41}, 195505 (2008).

\bibitem{Grisenti06}
R. E. Grisenti, R. A. Costa Fraga, N. Petridis, R. D\"orner, and J. Deppe, Europhys. Lett. \textbf{73}, 540 (2006).

\bibitem{Hemberg00}
O. Hemberg, B. A. M. Hansson, M. Berglund, and H. M. Hertz, J. Appl. Phys. \textbf{88}, 5421 (2000).

\bibitem{Bagnoud10}
V. Bagnoud, B. Aurand, A. Blazevic, S. Borneis, C. Bruske, B. Ecker, U. Eisenbarth, J. Fils, A. Frank, E. Gaul, S. Goette, C. Haefner, T. Hahn, K. Harres, H.-M. Heuck, D. Hochhaus, D. H. H. Hoffmann, D. Javorkov\'a, H.-J. Kluge, T. Kuehl, S. Kunzer, M. Kreutz, T. Merz-Mantwill, P. Neumayer, E. Onkels, D. Reemts, O. Rosmej, M. Roth, T. Stoehlker, A. Tauschwitz, B. Zielbauer, D. Zimmer, and K. Witte, Appl. Phys. B \textbf{100}, 137 (2010).


\end{thebibliography}

\end{document}